\documentclass[epj]{svjour}
\usepackage{graphics}

\begin{document}

\title{Scalar-isovector \boldmath $\mathsf{K\bar K}$ production close to threshold}

\author{A.~Dzyuba\inst{1,2} \and 
        V.~Kleber\inst{3}\thanks{\emph{Present address:}
            Physikalisches Institut, Universit\"at Bonn, Nuss\-allee 12,
            53115 Bonn, Germany} \and 
        M.~B\"uscher\inst{2}\thanks{\emph{m.buescher@fz-juelich.de}} \and 
        V.P.~Chernyshev\inst{4}\thanks{deceased}\and
        S.~Dymov\inst{5}\and
        P.~Fedorets\inst{2,4} \and
        V.~Grishina\inst{6} \and
        C.~Hanhart\inst{2} \and
        M.~Hartmann\inst{2} \and
        V.~Hejny\inst{2} \and
        L.~Kondratyuk\inst{5} \and
        V.~Koptev\inst{1} \and
        P.~Kulessa\inst{7} \and
        Y.~Maeda\inst{2,3}\thanks{\emph{Present address:}
            Research Center for Nuclear Physics, Osaka University, 
            Ibaraki, Osaka 567-0047, Japan} \and
        T.~Mersmann\inst{8} \and
        S.~Mikirtychyants\inst{1} \and
        M.~Nekipelov\inst{1,2} \and
        D.~Prasuhn\inst{2} \and
        R.~Schleichert\inst{2} \and
        A.~Sibirtsev\inst{2,9} \and
        H.J.~Stein\inst{2} \and
        H.~Str\"oher\inst{2} \and 
        I.~Zychor\inst{10}
}
\institute{
High Energy Physics Department, Petersburg Nuclear
  Physics Institute, 188350 Gatchina, Russia
\and
Institut f\"ur Kernphysik, Forschungszentrum J\"ulich,
  52425 J\"ulich, Germany
\and
Institut f\"ur Kernphysik, Universit\"at zu K\"oln, 50937 K\"oln, Germany
\and
Institute for Theoretical and Experimental Physics,
  Bolshaya Cheremushkinskaya 25, 117218 Moscow, Russia
\and
Laboratory of Nuclear Problems, Joint Institute for Nuclear Research, 141980
Dubna, Russia
\and
Institute for Nuclear Research, 60th October Anniversary Prospect 7A,
  117312 Moscow, Russia
\and
Institute of Nuclear Physics, Radzikowskiego 152, 31342 Cracow, Poland
\and
Institut f\"ur Kernphysik, Universit\"at M\"unster, 48149 M\"unster, Germany
\and
Helmholtz-Institut f\"ur Strahlen- und Kernphysik (Theorie), 
Universit\"at Bonn, Nu\ss allee 14-16, 53115 Bonn, Germany
\and
The Andrzej So{\l}tan Institute for Nuclear Studies, 05400 \'Swierk, Poland
}

\date{Received: date / Revised version: date}
%
\abstract{The reaction $pp {\to} d K^+ \bar{K^0}$ has been
  investigated at excess energies $Q{=47.4}$ and 104.7~MeV above the
  $K^+\bar{K^0}$ threshold at COSY-J\"ulich. Coincident $dK^+$ pairs
  were detected with the ANKE spectrometer, and subsequently
  $\sim2000$ events with a missing $\bar{K^0}$ invariant-mass were
  identified, which fully populate the Dalitz plot.  The joint
  analysis of invariant-mass and angular distributions reveals
  $s$-wave dominance between the two kaons, in conjunction with a
  $p$-wave between the deuteron and the kaon pair, {\em i.e.\/} $K\bar
  K$ production via the $a_0^+(980)$ channel.  Integration of the
  differential distributions yields total cross sections of $\sigma(pp
  {\to} d K^+ \bar{K^0}) {=} (38 {\pm} 2_\mathrm{stat} {\pm}
  14_\mathrm{syst})$\,nb and $(190 {\pm} 4_\mathrm{stat} {\pm}
  39_\mathrm{syst})$\,nb for the low and high $Q$ value,
  respectively.
} 

\maketitle

\section{Introduction}
\label{intro}

QCD is the fundamental theory of Strong Interactions. How quarks and
gluons are bound into hadrons is as yet an unsolved strong-coupling
problem. Though QCD can be treated explicitly in this regime using
lattice techniques~\cite{lattice}, these are not in a state to
make quantitative statements about the light scalar mesons.
Alternatively, QCD-inspired models, which employ effective degrees of
freedom, can be used. The constituent quark model is one of the most
successful in this respect (see {\em e.g.\/} Ref.~\cite{Morgan}). This
approach inherently treats the lightest scalar resonances
$a_0/f_0$(980) as conventional $q\bar{q}$ states.

Experimentally more states with quantum numbers $J^P{=}0^+$ have been
identified than would fit into a single SU(3) scalar nonet: the
$f_0(600)$ (or $\sigma$), $f_0(980)$, $f_0(1370)$, $f_0(1500)$ and
$f_0(1710)$ with $I{=}0$, the putative $\kappa(800)$ and the
$K^*(1430)$ ($I{=}1/2$), as well as the $a_0(980)$ and $a_0(1450)$
($I{=}1$)~\cite{PDG}. Consequently, the $a_0/f_0$(980) have been
associated with crypto-exotic states like $K\bar{K}$
mole\-cules~\cite{Wein} or compact $qq$-$\bar{q}\bar{q}$
states~\cite{Achasov}. It has even been suggested that a complete
nonet of four-quark states might exist with masses below
1.0~GeV/c$^2$~\cite{4q_nonet}.

The first clear observation of the isovector $a_0(980)$ resonance was
achieved in $K^-p$ interactions~\cite{Gay}. It has also been seen in
$p\bar{p}$ annihilations~\cite{Abele}, in $\pi^-p$
collisions~\cite{Teig}, and in $\gamma\gamma$
interactions~\cite{Achard}. Experiments on radiative
$\phi$-decays~\cite{Achas,Alois} have been analysed in terms of
$a_0/f_0$ production in the decay chain $\phi\to \gamma{a_0}/ f_0\to
\gamma\pi^0\eta/\pi^0\pi^0$. In $pp$ collisions the $a_0(980)$
resonance has been measured at $p_p=450$~GeV/c \textit{via}
$f_1(1285)\to a_0^{\pm}\pi^{\mp}$ decays~\cite{Barb} and in inclusive
measurements of the $pp\to dX^+$ reaction at $p_p=3.8$, 4.5, and
6.3~GeV/c~\cite{Abol}.

Despite these many experimental investigations, basic properties and
even the nature of the $a_0(980)$ resonance are still far from being
established (see {\em e.g.\/} Refs.~\cite{Baru:2003qq,Baru:2004xg}).
The Particle Data Group quotes a mass of
$m_{a_0}=(984.7\pm1.2)$~MeV/c$^2$ and a width of
$\Gamma_{a_0}=(50-100)$~MeV/c$^2$~\cite{PDG}. The main decay channels,
$\pi\eta$ and $K\bar K$, are denoted as ``dominant'' and ``seen'',
respectively. The corresponding coupling constants $g_{\pi\eta}$ and
$g_{K\bar K}$ differ significantly for the different data sets and
analyses~\cite{Baru:2004xg}.

Therefore, an experimental programme has been started at the Cooler
Synchrotron COSY-J\"ulich~\cite{Maier} aimed at exclusive data on the
$a_0/f_0$(980) production from $pp$, $pn$, $pd$ and $dd$ interactions
at energies close to the $K\bar{K}$ threshold~\cite{a0}. The final
goal of these investigations is the extraction of the $a_0/f_0$-mixing
amplitude, a quantity which is believed to shed light on the nature of
these resonances~\cite{Hanhart,Achas79}. As a first step the reactions
$pp\to d K^+\bar{K^0}$ \cite{a+_PRL} and $pp\to d\pi^+\eta$
\cite{Fedorets:2005bw} have been measured in parallel at the ANKE
spectrometer~\cite{ank} for $T_p$=2.65~GeV, corresponding to an excess
energy of $Q$=47.4~MeV with respect to the $K^+\bar{K^0}$ threshold.
The data for the strangeness decay channel --- which are almost
background free --- indicate that more than 80\% of the kaon pairs are
produced in a relative $s$-wave, corresponding to the $a_0^+$
channel~\cite{a+_PRL}.  On the other hand, the $\pi^+\eta$ signal sits
on top of a strong but smooth background of multi-pion production
which, together with the small acceptance of ANKE for this channel,
makes the interpretation of the $a_0^+$ signal
model-dependent~\cite{Fedorets:2005bw}. However, the obtained
branching ratio $\sigma(pp{\to}d(K^+\bar K^0)_{s\mathrm{-wave}})/
\sigma(pp{\to}da_0^+{\to}d\pi^+\eta) = 0.029\pm 0.008_{\mathrm{stat}}
\pm 0.009_{\mathrm{sys}}$~\cite{Fedorets:2005bw} is in line with
values from literature~\cite{PDG}.

In this paper we report on a refined analysis of the $pp\to d
K^+\bar{K^0}$ data at $T_p$=2.65~GeV as well as on new results from a
second measurement at higher beam energy ($T_p=2.83$~GeV, corresponding
to $Q=104.7$ MeV). The procedures for event identification and
acceptance correction at the lower energy have been described in our
previous publication~\cite{a+_PRL}.\\

\section{Measurement of \boldmath$\mathsf{pp\to d K^+\bar{K^0}}$ 
  events with ANKE}
\label{sec:experiment}

\subsection{Experimental setup}
\label{sec:anke}

ANKE is a magnetic spectrometer located in one of the straight
sections of COSY and comprises three dipole magnets, D1 --
D3~\cite{ank}.  D1 deflects the circulating COSY beam onto the target
in front of D2, and D3 bends it back into the nominal orbit. The
C-shaped spectrometer dipole D2 separates forward-going reaction
products from the COSY beam and allows one to determine their emission
angles and momenta. The angular acceptance of ANKE covers
$|\vartheta_{\mathrm h}| \le 10^{\circ}$ horizontally and
$|\vartheta_{\mathrm v}|\le 3^{\circ}$ vertically for the detected
deuterons ($p_d>1300$~MeV/c), and $|\vartheta_h| \le 12^{\circ}$ and
$|\vartheta_v| \le 3.5^{\circ}$ for the $K^+$ mesons.

A cluster-jet target~\cite{tar} of hydrogen molecules, placed between
D1 and D2, has been used, providing areal densities of up to ${\sim} 5
{\times} 10^{14}$ cm$^{-2}$.  The luminosity has been measured with
high statistical accuracy using $pp$ elastic scattering, recorded
simultaneously with the $dK^+$ data. Protons with $\vartheta=
5.5^\circ - 9^\circ$ have been selected, since the ANKE acceptance
changes smoothly in this angular range and the elastic peak is easily
distinguished from background in the momentum distribution. The
average luminosity during the measurements with up to $\sim 4\times
10^{10}$ stored protons in the COSY ring has been determined as
$L=(1.7 {\pm} 0.4_\mathrm{syst}) {\times}
10^{31}$\,s$^{-1}$\,cm$^{-2}$, corresponding to an integrated value of
$L_\mathrm{int}=7.5$\,pb$^{-1}$.

\subsection{Event selection for \boldmath $\mathsf{Q=104.7}$ MeV}
\label{sec:id}
Two charged particles, $K^+$ and $d$, have been detected in
coincidence. Positively charged kaons are identified in the side
detection system (SD)~\cite{ank,K_NIM} of ANKE by a time-of-flight
(TOF) measurement. The TOF-start counters, consisting of one layer of
23 scintillation counters, have been mounted next to the large exit
window of the vacuum chamber in D2. Kaons from $a_0^+$ decay with
momenta $p_{K^+}=390 - 625$ MeV/c have been stopped in range
telescopes, located along the focal surface of D2. These telescopes
comprise TOF-stop counters and provide additional kaon-{\em
  vs}-background discrimination by means of energy-loss ($\Delta E$)
measurements~\cite{K_NIM}. At $T_p=2.83$~GeV, kaons with $p_{K^+}=(625
- 1000)$ MeV/c have been detected in a different part of the SD,
consisting of one layer of 6 scintillation counters for TOF-stop
(``sidewall counters'').  Two multi-wire proportional chambers (MWPCs)
positioned between the TOF-start and -stop counters allow one to
deduce the ejectile momenta and to suppress background from secondary
scattering~\cite{a+_PRL,Fedorets:2005bw}.

\begin{figure}[htb]
 \begin{center}
  \scalebox{0.9}[0.85]
    {\resizebox{\linewidth}{!}{\includegraphics{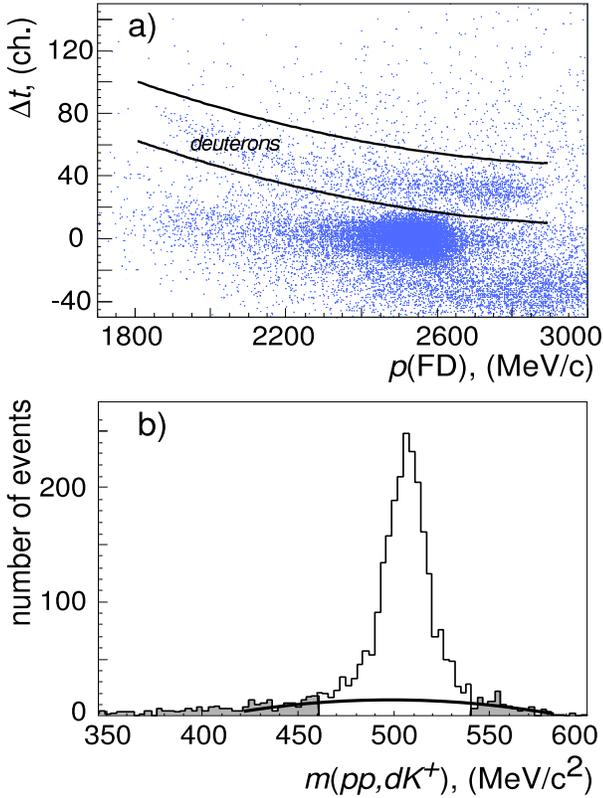}}}
  \vspace*{-4mm}
 \end{center}
    \caption{a) Time difference between the fast forward-going
      particles in layer 1 of the FD scintillators and the
      $K^+$-mesons {\em vs.\/} the momentum of the forward particle.
      The lines indicate the selection for deuteron identification.
      b) Missing-mass $m(pp,dK^+)$ distribution of the $pp {\to}
      dK^+X$ events. The shaded areas indicate the events used for
      background subtraction, the solid line shows the background
      distribution under the $\bar K^0$ peak obtained from a
      polynomial fit.}  \label{fig:tof_p}
\end{figure}

Fast particles produced in coincidence with the $K^+$ mesons as well
as elastically scattered protons have been detected in the ANKE
forward-detection system (FD)~\cite{Dymov:2005nc} containing two
layers of scintillation counters for TOF and $\Delta E$ measurements.
In addition there are three MWPCs, each with two sensitive planes,
exploited for momentum reconstruction and background
suppression~\cite{a+_PRL,ank}.  Two bands of protons and deuterons are
distinguished in the time difference between the detection of a
$K^+$-meson in one of the TOF-stop counters of the SD and a particle
in the FD as a function of the FD particle momentum, see
Fig.~\ref{fig:tof_p}a. The deuterons are selected with the cut
indicated by the lines, plus the energy-loss information from the FD
scintillation counters.  In Fig.~\ref{fig:tof_p}b the missing-mass
distribution $m(pp,dK^+)$ for the selected $pp {\to} d K^+ X$ events
is presented. The missing particle $X$ must be a $\bar K^0$, due to
charge and strangeness conservation. The measured $dK^+$ missing-mass
distribution peaks around $m{=}m_{\bar K^0}$, reflecting the clean
particle identification at ANKE.

About 2300 events are accepted as $dK^+\bar{K^0}$ candidates for
further analysis (unshaded peak area of the histogram in
Fig.~\ref{fig:tof_p}b)). The remaining background from misidentified
particles is ($13 {\pm} 2$)\,\%. The shape of this background in the
differential spectra discussed below has been determined and
subsequently subtracted by selecting events outside the $\bar K^0$
peak (shaded areas in Fig.~\ref{fig:tof_p}b).

The $K^+$ tracking efficiency in the side MWPCs and the efficiency of
the $\Delta E$ cut have been determined using simultaneously recorded
$pp {\to} p K^+\Lambda$ events, which, due to the significantly larger
cross section, can be identified by TOF criteria and a momentum cut
for protons in the FD only. The efficiency of the track reconstruction
varies from 96\% for the telescopes to 76\% for the sidewall counters.
The efficiency of the $\Delta E$ cut has been determined for each
telescope, with an average value of 53\%, and for the sidewall
counters, ranging from 87\% to 74\%.  The efficiency of the FD $\Delta
E$ criterion for deuterons has been deduced from the number of $\bar
K^0$ events in the peak of Fig.~\ref{fig:tof_p}b before and after this
cut. The efficiency of the FD scintillators and all TOF criteria is
larger than 99\%. The data have been corrected for all efficiencies on
an event-by-event basis.

\subsection{Kinematic fit}
\label{sec:kinfit}

A kinematic fit has been carried out to improve the in\-variant-mass
and angular resolutions. This fit shifts the measured $dK^+$ missing
mass (Fig.~\ref{fig:tof_p}b) to the nominal value of
$m_{\bar{K}^0}=497.6\,$MeV/c$^2$ on an event-by-event basis, varying
the momentum components of the detected $K^+$ and $d$ within their
resolutions. As a result of the fit, the deuteron missing-mass ({\em
  i.e.\/} the invariant ${K^+\bar{K}^0}$ mass) resolution improves
from $\delta m_{K^+\bar{K}^0}= (35-3)\,$MeV/c$^2$ over the range (991
-- 1096) MeV/$c^2$ to $\delta m_{K^+\bar{K}^0}<10\,$MeV/c$^2$ in the
full range with minimum values of $\sim 3\,$MeV/c$^2$ at the kinematic
limits.  Due to the fact that the $p_z$ resolution ($z$ being the beam
direction) for deuterons is approximately a factor five worse than for
all other variables, the fit procedure does not significantly improve
the $K^+$ missing-mass and angular resolutions: $\delta
m_{d\bar{K}^0}\sim$ 5 MeV/c$^2$ in the full range, $\delta
[\cos(\theta)]\sim 0.2$ for all angular spectra.

The same fit procedure has also been applied to the previously
published data at $T_p=$2.65 GeV~\cite{a+_PRL}, and improves $\delta
m_{K^+\bar{K}^0}$ from (8 -- 1) MeV/c$^2$ over the range (991 -- 1038)
MeV/c$^2$ to $\delta m_{K^+\bar{K}^0}<3\,$MeV/c$^2$ in the full mass
range.

\subsection{Non-acceptance-corrected Dalitz plot}
\label{sec:dalitz}

\begin{figure}[h]
  \begin{center}
  \vspace*{-3mm}
  \scalebox{1.0}[0.97]
  {\resizebox{\linewidth}{!}{\includegraphics{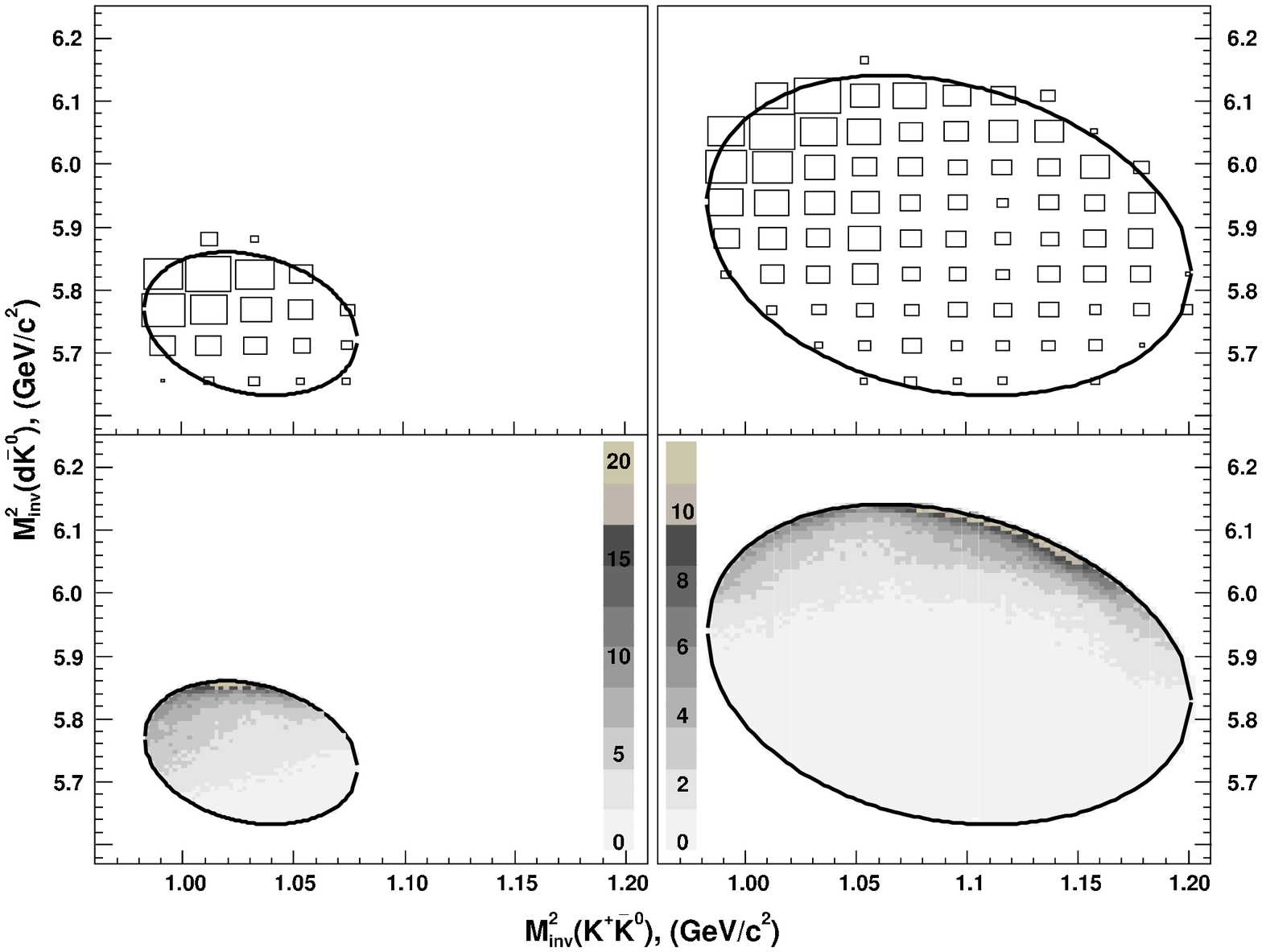}}}
  \vspace*{-7mm}
  \end{center}
  \caption{Upper: Dalitz plot of the events from the reaction
    $pp{\to}dK^+{\bar K^0}$ at $Q=47.4$ (left) and 104.7~MeV (right).
    The data are not background subtracted and not corrected for the
    ANKE acceptance and the detection efficiencies, and are binned
    with cell size $21\, \mathrm{MeV}^2/c^4 \times 57\,
    \mathrm{MeV}^2/c^4 $ The lines denote the kinematically allowed
    region. Lower: Simulated Dalitz-plots inside the ANKE acceptance
    for phase-space distributed events ({\em i.e.\/} configuration
    $[(K\bar{K})_s d]_s$, cf.\ Sect.~\ref{sec:acceptance}).}
\label{fig:dalitz}
\end{figure}

Figure \ref{fig:dalitz} shows the distribution of the kinematically
fitted $dK^+\bar{K^0}$ events in the Dalitz plot for both $Q$ values.
It is observed that the kinematically allowed region is fully covered
by the ANKE acceptance.  For comparison the simulated population of
the Dalitz plot is also shown for the case of phase-space-distributed
events.  The total ANKE acceptance for these $dK^+\bar{K^0}$ events is
2.1\% at $Q=47.4$~MeV and 0.8\% at 104.7~MeV.  Due to the limited
number of counts we present in the following only one-dimensional
distributions. These also have the advantage of carrying additional
information about the transition matrix, as shown in
Sect.~\ref{sec:acceptance}.

\subsection{Acceptance correction}
\label{sec:acceptance}
In comparison to the data at $T_p=2.65$~GeV~\cite{a+_PRL}, the excess
energy for the higher beam energy is approximitely twice as large. As
a consequence, the method of model-independent acceptance correction
(using a five-dimen\-sional acceptance matrix) can no longer be used,
since the number of zero elements in the acceptance matrix becomes too
large.  An alternative method has been developed which is described as
follows.

In the close-to-threshold regime only a limited number of final states
contribute. For the data ana\-ly\-sis we have restricted ourselves to
the lowest allowed partial waves, {\em i.e.\/} $s$-wave in the
$K\bar{K}$ system accompanied by a $p$-wave of the deuteron with
respect to the meson pair ($a_0^+(980)$-channel) and $p$-wave
$K\bar{K}$ production with an $s$-wave deu\-teron (non-resonant
channel). In the following we denote these two configurations by
$[(K\bar{K})_s d]_p$ and $[(K\bar{K})_p d]_s$.  It has been shown that
the lower energy data can be described by this {\em ansatz} for the
$dK^+\bar{K}^0$ final state~\cite{a+_PRL}, where the square of the
spin-averaged transition matrix element can be written as:
\begin{eqnarray}
\lefteqn{\vert\bar{\mathcal{M}}\vert ^{2} = C_{0}^{q}q^{2} + C_{0}^{k}k^{2} + 
 C_{1}(\hat{\vec p} \cdot \vec{k})^{2}} \nonumber\\
  && + C_{2}(\hat{\vec p} \cdot \vec{q})^{2} + C_{3}(\vec{k} \cdot \vec{q}) + 
          C_{4}(\hat{\vec p} \cdot \vec{k})(\hat{\vec p} \cdot \vec{q})\ .
\label{eq:m2}
\end{eqnarray}  

Here $\vec{k}$ is the deuteron momentum in the overall CMS, $\vec{q}$
denotes the $K^{+}$ momentum in the $K\bar{K}$ system, and $\hat{\vec
  p}$ is the unit vector of the beam momentum. Only $K\bar{K}$
$p$-waves contribute to $C_{0}^{q}$ and $C_{2}$, only $K\bar{K}$
$s$-waves to $C_{0}^{k}$ and $C_{1}$, and only $s$-$p$ interference
terms to $C_{3}$ and $C_{4}$.  The coefficients $C_i$ can be
determined from the data by fitting Eq.(\ref{eq:m2}) to the measured
${\mathrm d}\sigma/{\mathrm d}m_{K\bar K}$ and ${\mathrm
  d}\sigma/{\mathrm d}m_{dK}$ as well as to the angular distributions
${\mathrm d}\sigma/{\mathrm d}[\cos{(\vec{p}\vec{k})}]$, ${\mathrm
  d}\sigma/ {\mathrm d}[\cos{(\vec{p}\vec{q})}]$, ${\mathrm
  d}\sigma/{\mathrm d}[\cos{(\vec{k}\vec{q})}]$ and ${\mathrm
  d}\sigma/{\mathrm d}[\cos{(\vec{p}\vec{t})}]$~\cite{Hanhart}
($\vec{t}$ represents the $K^+$ momentum measured in the overall CMS).
It should be noted that a fit to the two-dimensional Dalitz plot does
not provide additional information about the transition matrix, but
would only yield three linear combinations of two of the coefficients
$C_i$~\cite{Hanhart}.

\begin{figure}[htb]
  \scalebox{1.0}[0.9]
   {\resizebox{\linewidth}{!}{\includegraphics{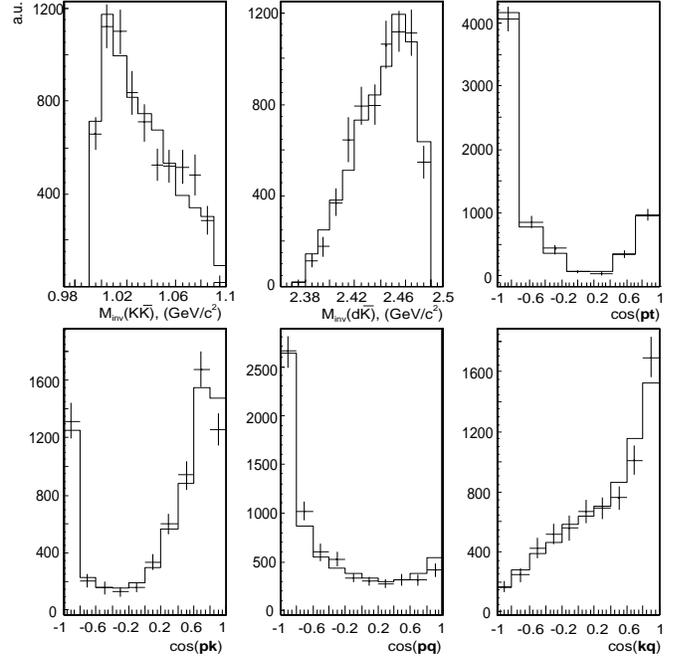}}}
  \vspace*{-3mm}
  \caption{Best fit to non-acceptance-corrected data at $T_p=2.83$
  GeV.}
\label{fig:dcs_2.83_wo_acceptance_correction}
\end{figure}

$\vert\bar{\mathcal{M}}\vert ^{2}$ gives the production probability of
an event with certain kinematic parameters $\vec{k}$ and $\vec{q}$
relative to $\hat{\vec p}$. The corresponding differential acceptance
of the spectrometer $\alpha(\vec{k},\vec{q},\hat{\vec p})$ does not
depend on the values of $C_i$, and can be determined using a large
sample of simulated events, covering full phase space, which are
tracked through a GEANT model of the setup~\cite{ANKE_GEANT}. Using
the coefficients from Ref.~\cite{a+_PRL} as starting parameters, the
simulations were carried out for different sets of the $C_i$, leading
to differential distributions convoluted with the acceptance. For each
choice of the $C_i$, the $\chi ^2$ values have been calculated for the
difference between simulated and measured distributions. Subsequently,
the coefficients which describe the experimental data best have been
determined by minimizing $\chi ^2$ with the MINUIT
package~\cite{James:1975dr}.  The best fit result of this procedure is
shown in Fig.~\ref{fig:dcs_2.83_wo_acceptance_correction} for two
invariant-mass and four angular distributions (cf.\
Table~\ref{tab:coeff} in Sect.~\ref{sec:diff} for numerical values).

\section{Cross sections of the reaction 
   \boldmath $\mathsf{pp\to dK^+\bar K^0}$ 
    at $\mathsf{Q=47.4}$ and 104.7 MeV}
\label{sec:results}

\subsection{Differential spectra}
\label{sec:diff}
\begin{figure}[htb]
  \scalebox{1.0}[0.9]
  {\resizebox{\linewidth}{!}{\includegraphics{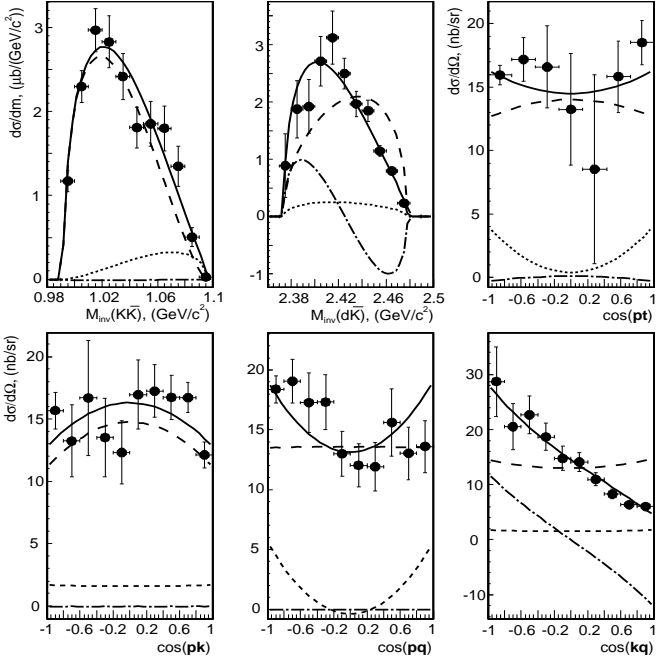}}}
  \caption{Angular and invariant mass distributions for $T_p=2.83$
    GeV. The dashed (dotted) line corresponds to $K\bar{K}$ production
    in a relative $s$- ($p$-)wave, the dash-dotted to the interference
    term, and the solid line is the sum of these contributions. The
    error bars show the statistical uncertainties only. The systematic
    uncertainty for each bin is smaller than 10\%; the overall
    uncertainty from the luminosity determination is given in
    Sect.~\ref{sec:anke}.}
\label{fig:dcs_2.83_w_acceptance_correction}
\end{figure}

With the best fit coefficients $C_i$ one can simulate corresponding
differential distributions at the target, track the events through the
setup, and thus determine one-dimen\-sional differential acceptances
for {\em e.g.\/} the two invariant masses and four angles of
Fig.~\ref{fig:dcs_2.83_wo_acceptance_correction}. Using these
acceptances, differential cross sections can be extracted from the
data, and these are shown in
Fig.~\ref{fig:dcs_2.83_w_acceptance_correction}.

\begin{figure}[htb]
  \scalebox{1.0}[0.9]
  {\resizebox{\linewidth}{!}{\includegraphics{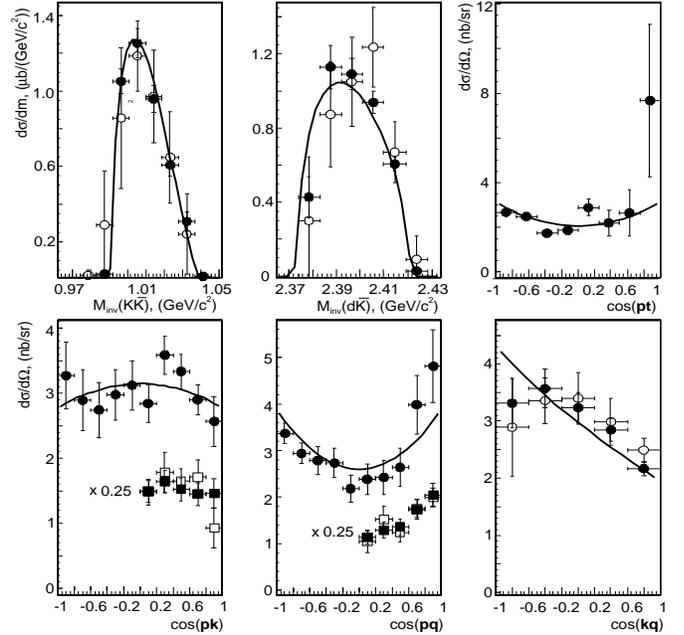}}}
  \caption{Same as Fig.~\ref{fig:dcs_2.83_w_acceptance_correction} for
    $T_p=2.65$ GeV but omitting the fitted partial-wave contributions.
    Full symbols denote the differential cross sections obtained by
    the method described in this paper; open symbols are the
    previously published model-independent results~\cite{a+_PRL} where
    the error bars include statistical and systematic uncertainties.
    The $\cos{(\vec{p}\vec{t})}$ spectrum has not been presented in
    Ref.~\cite{a+_PRL}. See text for further details.}
\label{fig:dcs_2.65_w_acceptance_correction}
\end{figure}

In order to verify the validity of the acceptance correction method
using the coefficients $C_i$, the same procedure has been applied to
the lower energy data.  The results are shown in
Fig.~\ref{fig:dcs_2.65_w_acceptance_correction} as solid dots and are
compared to our results published previously~\cite{a+_PRL}, analyzed
using the acceptance matrix method (open circles). For a $pp$ initial
state all distributions must be forward-backward symmetric relative to
the beam momentum; this feature has been exploited in
Ref.~\cite{a+_PRL}, where differential cross sections as functions of
$|\cos{(\vec{p}\vec{q})}|$ and $|\cos{(\vec{p}\vec{k})}|$ are
presented. These are shown in the lower left spectra of
Fig.~\ref{fig:dcs_2.65_w_acceptance_correction} together with the
mirrored distributions from the coefficient method (squares), each
scaled by 0.25 for better distinction. In all cases good agreement
between the model-independent matrix method~\cite{a+_PRL} and the {\em
  ansatz} discussed here is observed. Note that the matrix method did
not allow us to extract the $\cos{(\vec{p}\vec{t})}$ distribution
(upper right in the Figure) from the 2.65~GeV data which is now
possible with the coefficient {\em ansatz}.

The best-fit coefficients $C_i$ are presented in Table~\ref{tab:coeff}
for both beam energies. All coefficients are given in units of
$C_{0}^{k}$. This is due to the fact that $\vert\bar{\mathcal{M}}\vert
^{2}$ from Eq.(\ref{eq:m2}) is proportional to the differential cross
sections, thus leaving one parameter undetermined in the fit. The
errors of the $C_i$ are obtained by varying each coefficient (allowing
the others to change) such that the total $\chi^2$ increases by one.

\begin{table*}[htb]
  \caption{Quality and results of the fit using Eq.(\ref{eq:m2}). 
    For the definition of $N$ see Eq.~\ref{eq:phasa}.}
\begin{center}
\begin{tabular}{c | c c c c c c | c | c}
\hline
$Q$, MeV & $C_{0}^{k}$ & $C_{0}^{q}$ & $C_{1}$ & $C_{2}$ & $C_{3}$ & $C_{4}$ & $\chi^{2}$/ndf  & $N$, $\mu$b$\,$MeV$^{-2}$  \rule[-2mm]{0mm}{6mm}\\
\hline
\hline
 47.4  & 1 & $-0.34^{+0.26}_{-0.21}$ & $-0.14^{+0.14}_{-0.13}$ & $1.23^{+0.32}_{-0.32}$ & $-0.44^{+0.16}_{-0.16}$ & $-0.76^{+0.30}_{-0.33}$ & 1.38 & $26.6\pm 10.9$ \rule[-2mm]{0mm}{6mm}\\
 104.7 & 1 & $-0.07^{+0.14}_{-0.24}$ & $-0.22^{+0.12}_{-0.11}$ & $1.04^{+0.36}_{-0.19}$ & $-1.45^{+0.20}_{-0.12}$ &  $0.09^{+0.25}_{-0.55}$  & 1.10 & $13.5\pm 3.0$ \rule[-2mm]{0mm}{6mm}\\
\hline
\end{tabular}
\end{center}
\label{tab:coeff}
\end{table*}

The parameters $C_i$ from Eq.(\ref{eq:m2}) can be directly related to
the different partial waves~\cite{a+_PRL}.  Their contributions to the
various observables are shown in
Fig.~\ref{fig:dcs_2.83_w_acceptance_correction}.  The occurrence of
interference terms in the $\bar K^0d$ invariant mass distribution is
due to the choice of the kinematic variables, {\em i.e.\/} relative
momentum of the kaons and that of the deuteron with respect to the
kaon pair.  Consequently, there is no interference term in the
$K^+\bar K^0$ mass distribution. To get a distribution for the other
invariant mass that is free of interferences one needs to switch to
the $\bar K^0d$ relative momentum and the $K^+$ momentum.  Then,
however, there will be an interference term in the $K^+\bar K^0$ mass
distribution. The method how to construct the variable transformation
is described in detail in Ref.~\cite{Sibirtsev:2004kk}.

Our fit reveals a strong dominance of the $K\bar K$ $s$-wave
production rate ({\em i.e.\/} via the $a_0^+(980)$ channel) for both
beam energies: $(95\pm 4)$\% and $(89\pm 4)$\% at $T_p=2.65$ GeV and
$2.83$ GeV, respectively.


The quality of the fit clearly supports the {\em ansatz} to include
only the lowest partial waves in the data analysis. It should be noted
that the growth of the amplitudes due to the centrifugal-barrier
factor is taken care off by Eq.(\ref{eq:m2}).  An essential question
is to understand the variation in the parameters $C_3$ and $C_4$, see
Table~\ref{tab:coeff}. As outlined above, these parameters emerge
solely from an interference of the $[(K\bar K)_s d]_p$ with the
$[(K\bar K)_p d]_s$ partial waves.  Therefore, if there were a
significant phase motion in one of these groups ({\em e.g.\/} due to
the strong final state interaction in the $a_0$ channel), a variation
with energy especially of $C_3$ and $C_4$ is expected. This point
clearly calls for more theoretical investigations.

\subsection{Total cross sections}
\label{sec:total}
Knowing the coefficients $C_i$, and thus the initial differential
distributions (Figs.~\ref{fig:dcs_2.83_w_acceptance_correction} and
\ref{fig:dcs_2.65_w_acceptance_correction}), the total acceptance and
the total cross sections can be evaluated. For the higher energy, a
value of $\sigma(pp \to d K^+ \bar{K^0}) = (190 \pm 4_\mathrm{stat}
\pm 39_\mathrm{syst})$\,nb is obtained. At the lower energy the
extracted total cross section is in agreement with the previously
published value of $(38 \pm 2_\mathrm{stat}\pm
14_\mathrm{syst})$\,nb~\cite{a+_PRL}. In both cases the errors include
the statistical and systematic uncertainty from the luminosity
determination.

\begin{figure}[t]
  \scalebox{1.0}[0.9]
  {\resizebox{\linewidth}{!}{\includegraphics{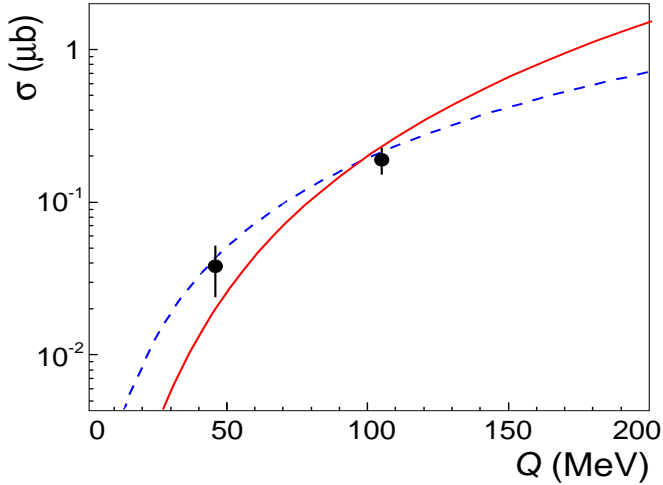}}}
  \caption{Total cross section of the $pp{\to}dK^+{\bar K^0}$ reaction
    as function of the excess energy $Q$. The solid line is the result
    from Eq.(\ref{eq:phasa}) with the squared transition amplitude
    given by Eq.(\ref{eq:transa}) and with
    $N{=}18$~$\mu$b$\,$MeV$^{-2}$.  The dashed line shows the energy
    dependence for three-body phase space. Note that the latter is
    forbidden by selection rules.}
\label{fig:a0anka3}
\end{figure}

Figure~\ref{fig:a0anka3} shows the measured total cross sections in
comparison with the expected $Q$ dependence of the cross section
calculated with the transition matrix element of Eq.(\ref{eq:m2}).
After an angular integration the total $pp\to dK^+{\bar K^0}$ cross
section is given by
\begin{eqnarray}
     \sigma = \frac{N}{2^6\pi^3\sqrt{s^2-4sm_p^2}}\ \ \!\!\!\!\!\!
   \int\limits_{4m_K^2}^{(\sqrt{s}-m_d)^2}\!\!\!\!\!\!
   \frac{k\, q}{\sqrt{s\,\, s_{KK}}} \, \, |{\tilde {\cal M}}|^2\,\, 
            {\mathrm d}s_{KK}\ ,
  \label{eq:phasa}
\end{eqnarray}
where $s$ and $s_{KK}$ are the squared invariant energies of the
initial $pp$ and final $K^+{\bar K^0}$ systems, respectively.  Here
$k$ and $q$ are defined as before and are given explicitly as
\begin{eqnarray}
   k^2&=&\frac{(s-s_{KK}-m_d^2)^2-4s_{KK}m_d^2}{4s}, \nonumber \\
   q^2&=&\frac{s_{KK}-4m_K^2}{4},\nonumber
\end{eqnarray}
where $m_d$ and $m_K$ are the deuteron and kaon masses, {\em i.e.\/}
we neglect the $K^+$ and ${\bar K^0}$ mass difference. The
angular-integrated squared transition amplitude $|{\tilde {\cal
    M}}|^2$ is given as
\begin{eqnarray}
   |{\tilde {\cal M}}|^2 = \left(C_0^q+\frac{1}{3}C_2\right)\, q^2 
+ \left(C_0^k+\frac{1}{3}C_1\right)\, k^2.
\label{eq:transa}
\end{eqnarray}

The normalization factor $N$ has been determined for both energies and
is quoted in Table~\ref{tab:coeff}.  The errors of $N$ include the
systematic uncertainties of the total cross sections.  Note that the
lines in Figs.~\ref{fig:dcs_2.83_w_acceptance_correction} and
\ref{fig:dcs_2.65_w_acceptance_correction} have been properly scaled
to the individual total cross sections.

For illustration we show by the dashed line in Fig.~\ref{fig:a0anka3}
the result of Eq.(\ref{eq:phasa}) with a constant matrix element.
This is a classical example when data can be well reproduced by a
simple phase-space consideration although such a description is
invalid for this particular reaction since selection rules do not
allow for a pure $s$-wave.

\section{Summary and conclusions}
\label{sec:conclusion}
Using the ANKE spectrometer at an internal target position of
COSY-J\"ulich, we have searched for scalar $K\bar K$ production in the
reaction $pp \to d K^+ \bar{K^0}$ at two excess energies $Q=47.4$ and
104.7~MeV. Due to the excellent $K^+$ identification at ANKE, the
detected events with coincident $K^+d$ pairs exhibit little
background. This can be subtracted using events outside the $\bar K^0$
missing-mass peak. After a kinematic fit to the $\bar K^0$ mass, the
invariant $K^+\bar{K^0}$ mass distribution has been obtained with an
unprecedented resolution of better than 3 (10) MeV/c$^2$ for 47.4
(104.7)~MeV.

Mass and angular distributions have been extracted from the data using
an {\em ansatz} for the transition matrix element that includes the
lowest allowed partial waves, {\em i.e.\/} an $s$-wave in the
$K\bar{K}$ system accompanied by a $p$-wave of the deuteron with
respect to the meson pair and $p$-wave $K\bar{K}$ production with an
$s$-wave deu\-teron.  All six coefficients that enter the
spin-averaged matrix element have been obtained by a fit to the
differential spectra. This fit reveals the dominance of $K\bar K$
production in a relative $s$-wave, $(95\pm 4)$\% and $(89\pm 4)$\% at
47.4 and 104.7~MeV, {\em i.e.\/} dominance of kaon-pair production via
the $a_0^+(980)$ channel.

The reaction $pp \to d K^+ \bar{K^0}$ has been subject of several
theoretical papers. For example, the authors of
Ref.~\cite{Oset:2001ep} point out that this reaction (and also $pp \to
d \pi^+ \eta$) is expected to be an additional source of information
about the scalar sector. They account for the interactions of the
mesons by using chiral unitary techniques, which dynamically generate
the $a_0^+(980)$ resonance. In Ref.~\cite{Grishina:2004rd} total cross
sections and differential spectra are calculated using a model in
which the reaction $pp {\to} d K^+\bar{K^0}$ is dominated by
intermediate $a_0^+(980)$ production.

As discussed in Sect.~\ref{sec:diff}, there is an energy dependence of
the parameters $C_3$ and $C_4$ in the transition matrix element
(Eq.(\ref{eq:m2})) that is not yet understood and needs further
theoretical study since it might indicate a final-state interaction in
the $a_0$ channel.


\section*{Acknowledgements}
This work has been carried out within the framework of the ANKE
Collaboration \cite{anke-collaboration} and supported by the COSY-FFE
program, the Deutsche Forschungs\-ge\-mein\-schaft (436 RUS 113/337, 444,
561, 768, 787), Russian Academy of Sciences (02-04-034, 02-04034,
02-18179a, 02-06518, 02-16349). We are grateful to the
COSY-accelerator crew for providing the proton beam above the nominal
accelerator energy.  Special thanks to C.~Wilkin for carefully reading
the manuscript and for conducting photomultiplier tests.

\end{document}